\documentclass[prb,showpacs,preprintnumbers,amsmath,amssymb,twocolumn,superscriptaddress,longbibliography]{revtex4-1}
\usepackage{graphicx}
\def\be{\begin{equation}}
\def\ee{\end{equation}}
\def\bea{\begin{eqnarray}}
\def\eea{\end{eqnarray}}

\begin{document}
\title{First order valence transition: Neutron diffraction, inelastic neutron scattering and x-ray absorption investigations on the double perovskite Ba$_{2}$PrRu$_{0.9}$Ir$_{0.1}$O$_{6}$}
\author{J. Sannigrahi}
\email{jhuma.sannigrahi@stfc.ac.uk}
\affiliation{ISIS Facility, Rutherford Appleton Laboratory, STFC, Chilton, Didcot, OX11 0QX, United Kingdom}
\author{D.T. Adroja}
\email{devashibhai.adroja@stfc.ac.uk}
\affiliation{ISIS Facility, Rutherford Appleton Laboratory, STFC, Chilton, Didcot, OX11 0QX, United Kingdom}
\affiliation{Highly Correlated Matter Research Group, Physics Department, University of Johannesburg, P.O. Box 524, Auckland Park 2006, South Africa}
\author{C. Ritter}
\affiliation{Institut Laue Langevin, 71 Rue des Martyrs, 38042 Grenoble, France}
\author{W. Kockelmann}
\affiliation{ISIS Facility, Rutherford Appleton Laboratory, STFC, Chilton, Didcot, OX11 0QX, United Kingdom}
\author{A.D. Hillier}
\affiliation{ISIS Facility, Rutherford Appleton Laboratory, STFC, Chilton, Didcot, OX11 0QX, United Kingdom}
\author{K.S. Knight}
\affiliation{ISIS Facility, Rutherford Appleton Laboratory, STFC, Chilton, Didcot, OX11 0QX, United Kingdom}
\author{A.T. Boothroyd}
\affiliation{Department of Physics, University of Oxford, Clarendon Laboratory, Parks Road, Oxford, OX1 3PU, United Kingdom}
\author{M. Wakeshima}
\affiliation{Hokkaido University School of Science, Chemistry Department, Sapporo, Hokkaido 0600810, Japan}
\author{Y. Hinatsu}
\affiliation{Hokkaido University School of Science, Chemistry Department, Sapporo, Hokkaido 0600810, Japan}
\author{F. Mosselmans}
\affiliation{Diamond Light Source, Harwell Science $\&$ Innovation Campus, Didcot, Oxon OX11 0DE, United Kingdom}
\author{S. Ramos}
\affiliation{Diamond Light Source, Harwell Science $\&$ Innovation Campus, Didcot, Oxon OX11 0DE, United Kingdom}
\affiliation{School of Physical Sciences, University of Kent, Canterbury, Kent, CT2 7NH, U.K.}

\date{\today}

\begin{abstract}
Bulk studies have revealed a first-order valence phase transition in Ba$_2$PrRu$_{1-x}$Ir$_x$O$_6$ ($0.10 \le x \le 0.25$), which is absent in the parent compounds with $x = 0$ (Pr$^{3+}$) and $x =1$ (Pr$^{4+}$), which exhibit antiferromagnetic order with transition temperatures $T_{\rm N} = 120$ and 72 K, respectively. In the present study, we have used magnetization, heat capacity, neutron diffraction, inelastic neutron scattering and x-ray absorption measurements to investigate the nature of the Pr ion in $x =0.1$. The magnetic susceptibility and heat capacity of $x =0.1$ show a clear sign of the first order valence phase transition below 175 K, where the Pr valence changes from 3+ to 4+. Neutron diffraction analysis reveals that $x =0.1$ crystallizes in a monoclinic structure with space group $P2_1/n$ at 300 K, but below 175 K two phases coexist, the monoclinic having the Pr ion in a 3+ valence state and a cubic one ($Fm\overline{3}m$) having the Pr ion in a 4+ valence state. Clear evidence of an antiferromagnetic ordering of the Pr and Ru moments is found in the monoclinic phase of the $x = 0.1$ compound below 110 K in the neutron diffraction measurements. Meanwhile the cubic phase remains paramagnetic  down to 2 K, a temperature below which heat capacity and susceptibility measurements reveal a ferromagnetic ordering. High energy inelastic neutron scattering data reveal well-defined high-energy magnetic excitations near 264 meV at temperatures below the valence transition. Low energy INS data show a broad magnetic excitation centred at 50 meV above the valence transition, but four well-defined magnetic excitations at 7 K. The high energy excitations are assigned to the Pr$^{4+}$ ions in the cubic phase and the low energy excitations to the Pr$^{3+}$ ions in the monoclinic phase. Further direct evidence of the Pr valence transition has been obtained from the x-ray absorption spectroscopy. The results on the $x =0.1$ compound are compared with those for $x=0$ and 1.
\end{abstract}
\maketitle
\section{Introduction}
In metallic solids there is a general correlation between the loss of magnetism and the decrease or increase of the unit-cell volume. This tendency is especially pronounced in Ce and Yb compounds, which have nearly degenerate electronic configurations, 4$f^n$ and 4$f^{n+1}$, where $n$ is the number of electrons in the 4$f$ shell.~\cite{freeman, yoshi, takabatake, hunley, malik, osborn} Due to the presence of two nearly degenerate states, these compounds display intermediate valence or non-integral valence behaviour. Generally the intermediate valence systems exhibit a gradual change in the valence state of the rare-earth ions with temperature, pressure and alloying.~\cite{lawrence, jayaraman, wada, zhang} However, there exist a handful of systems~\cite{felner, felner1, adroja} in which the valence transition is of first-order as a function of temperature or pressure while crystal symmetry remains the same below and above the phase transition. A classic example is Ce-metal, which shows a pressure induced first-order isostructural transition (at 8 kbar at 300 K) from a magnetic $\gamma $-phase to a non-magnetic $\beta $-phase.~\cite{koskenmaki}
\par
Very recently, it has been shown through magnetic susceptibility measurements that Ba$_2$PrRu$_{1-x}$Ir$_x$O$_6$ (with $0.1\le x \le 0.25$) compounds~\cite{wakeshima} also exhibit a temperature-induced valence transition, similar to those observed in CeNi$_{1-x}$Co$_x$Sn (Ref.~\onlinecite{adroja1}) and YbInCu$_4$ (Ref.~\onlinecite{sato}). Unlike in the Ce and Yb compounds, however, in which one valence state (Ce$^{3+}$/Yb$^{3+}$) is magnetic and the other (Ce$^{4+}$/Yb$^{2+}$) has zero magnetic moment,~\cite{felner, adroja} the valence phase transition in the Pr compounds is especially interesting as both Pr$^{3+}$ and Pr$^{4+}$ carry a magnetic moment.

Ba$_2$PrRu$_{1-x}$Ir$_x$O$_6$ compounds crystallize in the perovskite-type structure, with a small monoclinic distortion, and have a  1$:$1 ordered arrangement between Pr$^{3+}$ and Ru$^{5+}$ (or Ir$^{5+}$) over the six-coordinate $B$ sites of the well known perovskite ABO$_3$ structure.~\cite{wakeshima, li} The lattice parameters and unit cell volume of the Ru-rich side of Ba$_2$PrRu$_{1-x}$Ir$_x$O$_6$ decrease with increasing Ir concentration, while for the Ir-rich side, they are nearly constant.~\cite{li} The change in the lattice parameters and unit cell volume is attributed to the change in the oxidation state of the Pr ion.

\par
A previous neutron diffraction study on Ba$_2$PrRuO$_6$ revealed type-I antiferromagnetic ordering of Pr$^{3+}$ and Ru$^{5+}$ moments below $T_{\rm N} \simeq 120$ K.~\cite{izumi, parkinson} Zero-field muon-spin rotation ($\mu$SR) measurements on Ba$_2$PrRu$_{1-x}$Ir$_x$O$_6$ with $x = 0$, 0.1 and 1.0 revealed in addition a coexistence of two phases in $x=0.1$ below the valence phase transition ($T_{\rm V}\simeq 170$ K), one phase being magnetically ordered and the other paramagnetic.~\cite{adrian}

The aim of the present study was to obtain direct information on the valence phase transition, changes in the crystal structure as well as details of the magnetic structure below the ordering temperature. To this end, high resolution and high intensity neutron powder diffraction, as well as inelastic neutron scattering measurements, were performed on Ba$_2$PrRu$_{0.9}$Ir$_{0.1}$O$_6$. Here we shall report how the valence state transition transforms the ground state magnetic structure in this compound as well as the nature of the crystal field ground state of the Pr$^{4+}$ ion. For comparison we also present additional results on Ba$_2$PrRuO$_6$.

\section{Experimental Details}
Polycrystalline samples of Ba$_2$PrRu$_{0.9}$Ir$_{0.1}$O$_6$ were prepared following a solid state reaction route in air. A stoichiometric mixture of BaCO$_3$, RuO$_2$, Ir metal and Pr$_6$O$_{11}$ was initially mixed homogeneously in the stoichiometric ratio, pelletized and calcined at 900$^\circ$C for 48 h. The final sintering was performed at 1200$^\circ$C for 3 days with several intermediate processes of regrinding and repelletizing.~\cite{wakeshima}
\par
The magnetic susceptibility ($\chi$) and heat capacity ($C_p$) measurements were made on a Quantum Design physical properties measurement system (PPMS).  Neutron powder diffraction patterns for Ba$_2$PrRu$_{0.9}$Ir$_{0.1}$O$_6$ were recorded on the time-of-flight (TOF) High Resolution Powder Diffractometer (HRPD) at the ISIS Neutron and Muon Facility, UK. Data were collected in the temperature range 4 to 300 K. The sample was finely ground and was contained in a thin walled vanadium can. The analysis of HRPD data was carried out with the GSAS software,~\cite{GSAS} and guided by group-subgroup relationships of ordered double perovskites. The background was defined by a third order Chebychev polynomial in TOF  in the Rietveld refinements. Further neutron diffraction measurements to investigate the magnetic structure were performed on the high intensity diffractometer D20 at the Institut Laue-Langevin, Grenoble, France, with a constant wavelength of 2.41 \AA.

Inelastic neutron scattering (INS) measurements were carried out on the High Energy Transfer (HET) TOF spectrometer at the ISIS Facility with incident neutron energies ($E_{\rm i}$) of 900, 500, 200, 100, 60 and 15 meV at temperatures of 222 K (above $T_{\rm N}$) and 5 K (below $T_{\rm N}$). The powder sample with $x=0.1$ was mounted in a thin Al-foil envelope, which was cooled down to 5 K in a closed cycle refrigerator (CCR) in the presence of helium exchange gas.

The Pr L$_3$-edge x-ray absorption near-edge structure (XANES) of $x = 0$, 0.1 and 1 compounds was measured in transmission mode between 77 K and 300 K on the beamline B18, the Core EXAFS (Extended X-ray Absorption Fine Structure) Beamline, at the Diamond Light Source, UK, and at station 7.1 of the Synchrotron Radiation Source (SRS) at the Daresbury Laboratory, UK. Samples were prepared by grinding the polycrystalline material into a fine powder, mixing it with cellulose and pressing the mixture into pellets.

\begin{figure}[t]
\begin{center}
\includegraphics[width=0.99\columnwidth]{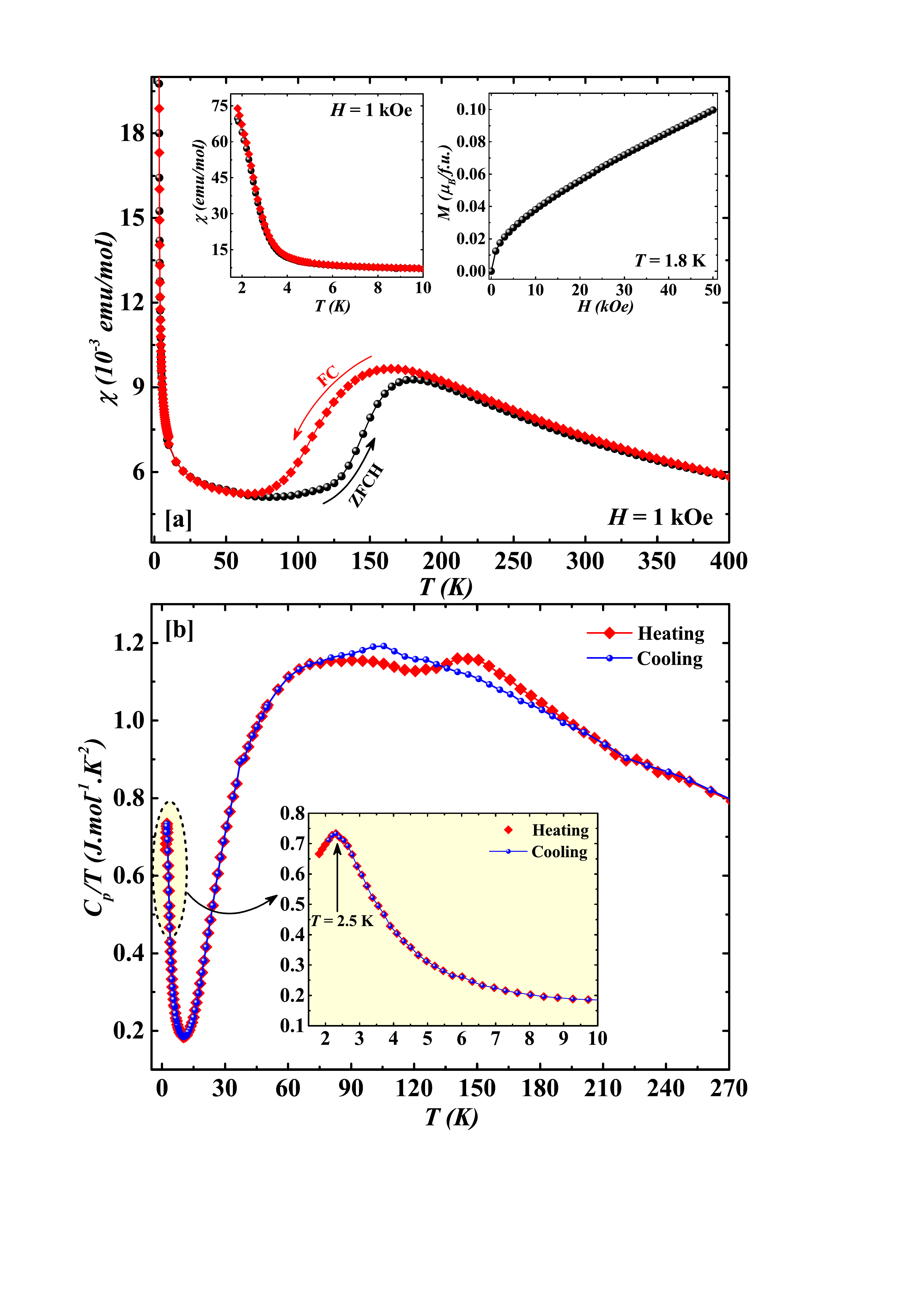}
\end{center}

\caption {(a) Temperature variation of the ZFC and FC susceptibility of Ba$_2$PrRu$_{0.9}$Ir$_{0.1}$O$_6$ in a measuring field of $H = 1$ kOe. The top left inset shows the susceptibility behavior at low temperature, indicating ferromagnetic ordering below 3 K, and the top right inset shows the magnetization isotherm at 1.8 K.  (b) Heat capacity data, plotted as $C_p/T$ vs $T$,  taken during heating and cooling cycles. The inset shows the low temperature behavior of $C_p/T$ vs $T$, which indicates a magnetic anomaly near 2.5 K.}
\label{MT}
\end{figure}

\section{Results and Discussion}
\subsection{Magnetization and heat capacity}
Figure~1(a) shows the temperature dependence of magnetic susceptibility of Ba$_2$PrRu$_{0.9}$Ir$_{0.1}$O$_6$ in an applied magnetic field of $H = 1$ kOe. This compound shows a distinct magnetic anomaly between 60 and 200 K. A drop in susceptibility is observed on cooling below 200 K for both zero-field-cooled (ZFC) and field-cooled (FC) cycles. Clear hysteresis between ZFC and FC data signifies the first order nature of this transition. This anomaly is ascribed to the valence phase transition of the Pr ions on the B-site.~\cite{wakeshima} The temperature range in which the ZFC and FC data separate from each other is the same as the temperature range in which two perovskite phases Ba$_2$Pr$^{3+}$(Ru,Ir)$^{5+}$O$_6$ and Ba$_2$Pr$^{4+}$(Ru,Ir)$^{4+}$O$_6$ coexist in accordance with x-ray diffraction measurements.~\cite{li} This shows that the valence state of the Pr ion changes from Pr$^{3+}$ (above 200 K) to Pr$^{4+}$ below 200 K. The inset (top right) of Fig.~1(a) shows the isothermal magnetization recorded at 1.8 K after ZFC. At low fields the magnetization shows ferromagnetic behaviour without hysteresis. The value of the saturated moment at 50 kOe is approximately 0.1 $\mu_{\rm B}/$f.u.
\par
The magnetic anomaly between 60 and 200 K is also clearly present in the temperature dependent heat capacity data shown in the main panel of Fig.~1(b). The cooling and heating curve start to separate below 200 K, consistent with the first order nature of the transition. Interestingly, we observed a maximum in the $C_p/T$ versus $T$ curve at $\approx$ 2.5 K (inset of Fig.~1(b)) which coincides with the sharp rise in the magnetic susceptibility below 3 K (inset of Fig.~1(a)) indicating ferromagnetic ordering. The origin of the ferromagnetic ordering is attributed to the ordering of the cubic phase in the two-phase separated states, as will be discussed in the Section~B2.

\begin{figure}
\includegraphics[width=8 cm,trim={100 0 300 100}]{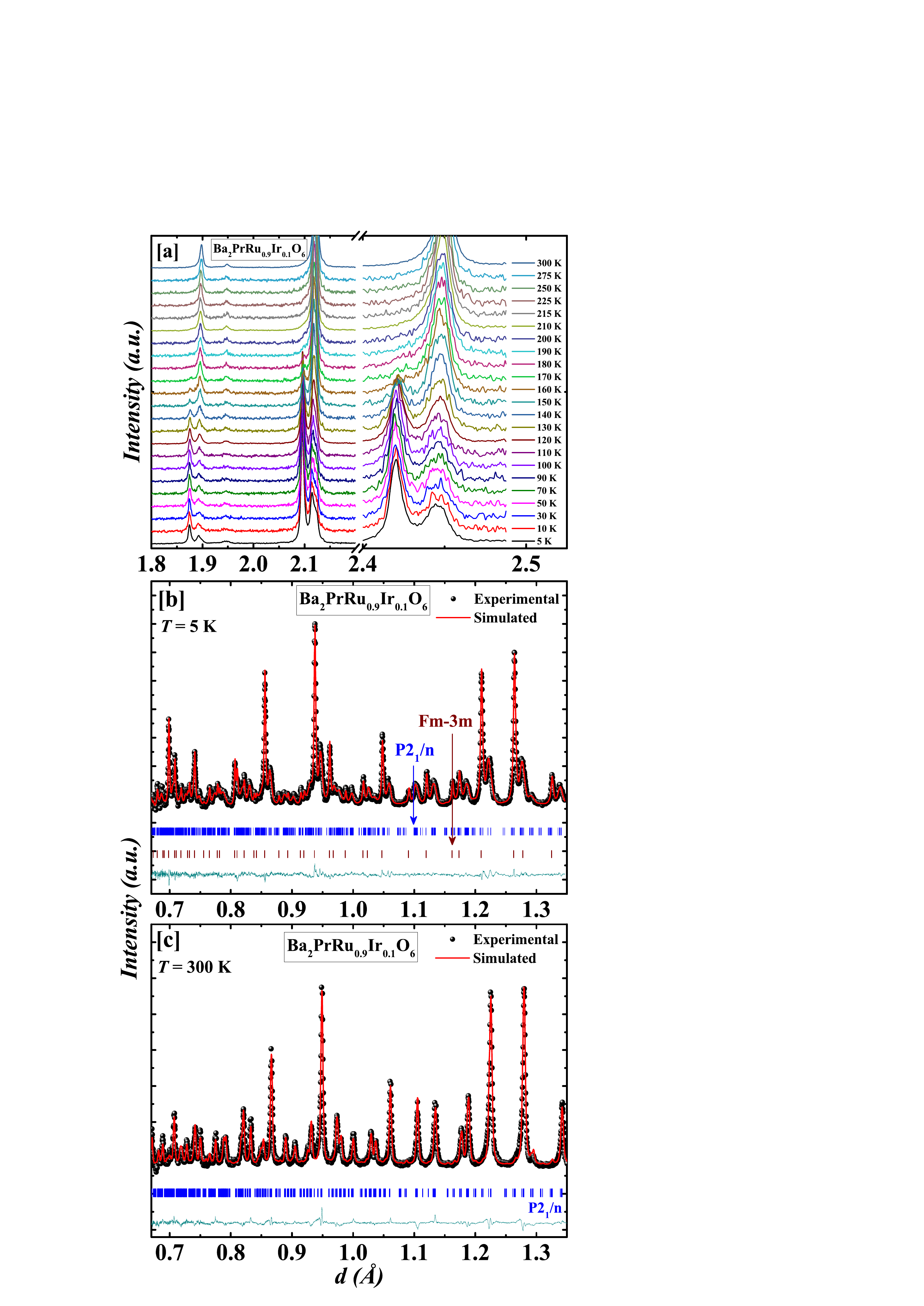}
\caption {(a) Powder neutron diffraction patterns of Ba$_2$PrRu$_{0.9}$Ir$_{0.1}$O$_6$ obtained from HRPD in the temperature range 5 to 300 K. (b) and (c) show Rietveld refinements of diffraction data measured at 5 K and 300 K, respectively.  Black circles indicate the experimental data points and red solid lines are the simulated curves. The Bragg peak positions of the monoclinic and cubic phases are indicated by blue and maroon vertical ticks, respectively, and the light-blue line shows the difference between the data and the simulated curve.}
\label{PND}
\end{figure}

\subsection{Neutron diffraction}
\subsubsection{Crystal structure}

\begin{figure}[t]
\includegraphics[width = 8 cm]{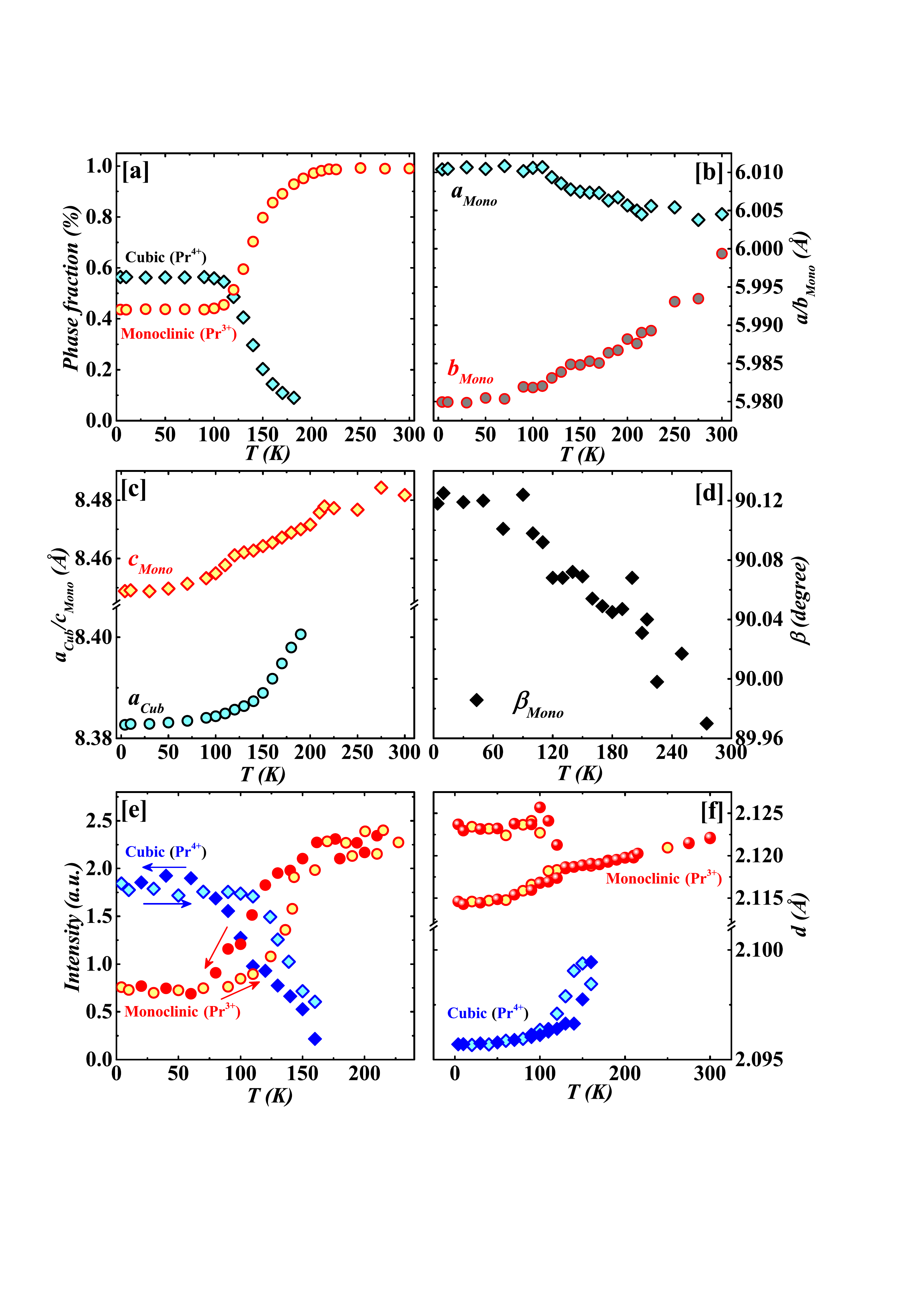}
\caption {Temperature dependence of structural parameters of Ba$_2$PrRu$_{0.9}$Ir$_{0.1}$O$_6$. (a) Fraction of monoclinic (space group $P2_{1}/n$) and cubic (space group $Fm\overline{3}m$) phases.  (b) Monoclinic lattice parameters $a_{\rm mono}$ and $b_{\rm mono}$. (c) Monoclinic lattice parameter $c_{\rm mono}$ and cubic lattice parameter $a_{\rm cub}$. (d) Temperature variation of the monoclinic angle $\beta_{\rm mono}$.  (e) Single peak analysis which shows the variation of the fraction of  monoclinic phase (containing Pr$^{3+}$) and cubic phase (containing Pr$^{4+}$) in warming and cooling measurements.  (f) The $d$-spacing of the peaks shown in (e).}
\label{PND1}
\end{figure}

\par
The temperature dependence of the powder neutron diffraction patterns of Ba$_2$PrRu$_{0.9}$Ir$_{0.1}$O$_6$ as obtained from HRPD is shown in Fig.~2(a), with heating cycles between 5 and 300 K. At 300 K, Fig.~2(c), the data can be refined using a single monoclinic phase with space group $P2_1/n$. As the temperature is lowered, extra peaks arise below about 170 K and persist down to 5 K, as shown in Fig.~2(a). Rietveld analysis shows that these extra peaks are due to the appearance of a new phase, which was identified as a cubic structure with space-group $Fm\overline{3}m$. A two-phase fit at 5 K is shown in Fig.~2(b). This new phase is therefore similar to the parent phase Ba$_2$PrIrO$_6$ (with Pr$^{4+}$), implying that the Pr ion in the cubic phase of Ba$_2$PrRu$_{0.9}$Ir$_{0.1}$O$_6$ is in the Pr$^{4+}$ state below $T_{\rm V} \simeq 170$ K, the valence phase transition temperature.~\cite{fu, winfred} It is to be noted that the Pr ions are in the Pr$^{3+}$ state in Ba$_2$PrRuO$_6$, and also in Ba$_2$PrRu$_{0.9}$Ir$_{0.1}$O$_6$ at 300 K. As the cubic phase Bragg reflections appear on cooling below $T_{\rm V}$ there is a concomitant decrease in the  reflections corresponding to the room temperature monoclinic phase, see Fig.~2(a).

In order to describe the crystal structures more quantitatively we performed detailed structural analysis via Rietveld refinement of the high resolution powder neutron diffractograms recorded between 5 and 300 K. Fig.~3(a) shows how the fraction of cubic and monoclinic phases changes with temperature. At 5 K the estimated volume faction is 58\% for the cubic phase and 42\% for the monoclinic phase. Above 125 K the fraction of the monoclinic phase sharply increases and the cubic phase becomes almost imperceptible above 170 K. The temperature dependence of the monoclinic lattice parameters and $\beta$ angle are shown in Figs.~3(b), (c) and (e). It can be seen that with decreasing cubic phase fraction the lattice parameters of the monoclinic phase start to change at about 100 K, such that at 300 K $a_{\rm mono}$ $\approx$ $b_{\rm mono}$ $\approx$ $c_{\rm mono}$/$\sqrt{2}$ and $\beta_{\rm mono}$ $\approx$ 90$^{\circ}$. This implies that the structure is nearly cubic at 300 K.

\par
The temperature dependence of the coexistence of the monoclinic and the cubic phases was investigated further by fitting the intensity of individual representative Bragg peaks from both phases during both heating and cooling cycles, see Figs.~3(e) and 3(f). This analysis clearly reveals the presence of a thermal hysteresis between 80 K and 170 K which confirms the first order nature of this transition.

\begin{figure}[t]
\includegraphics[width = 8 cm, trim={100 50 250 350}]{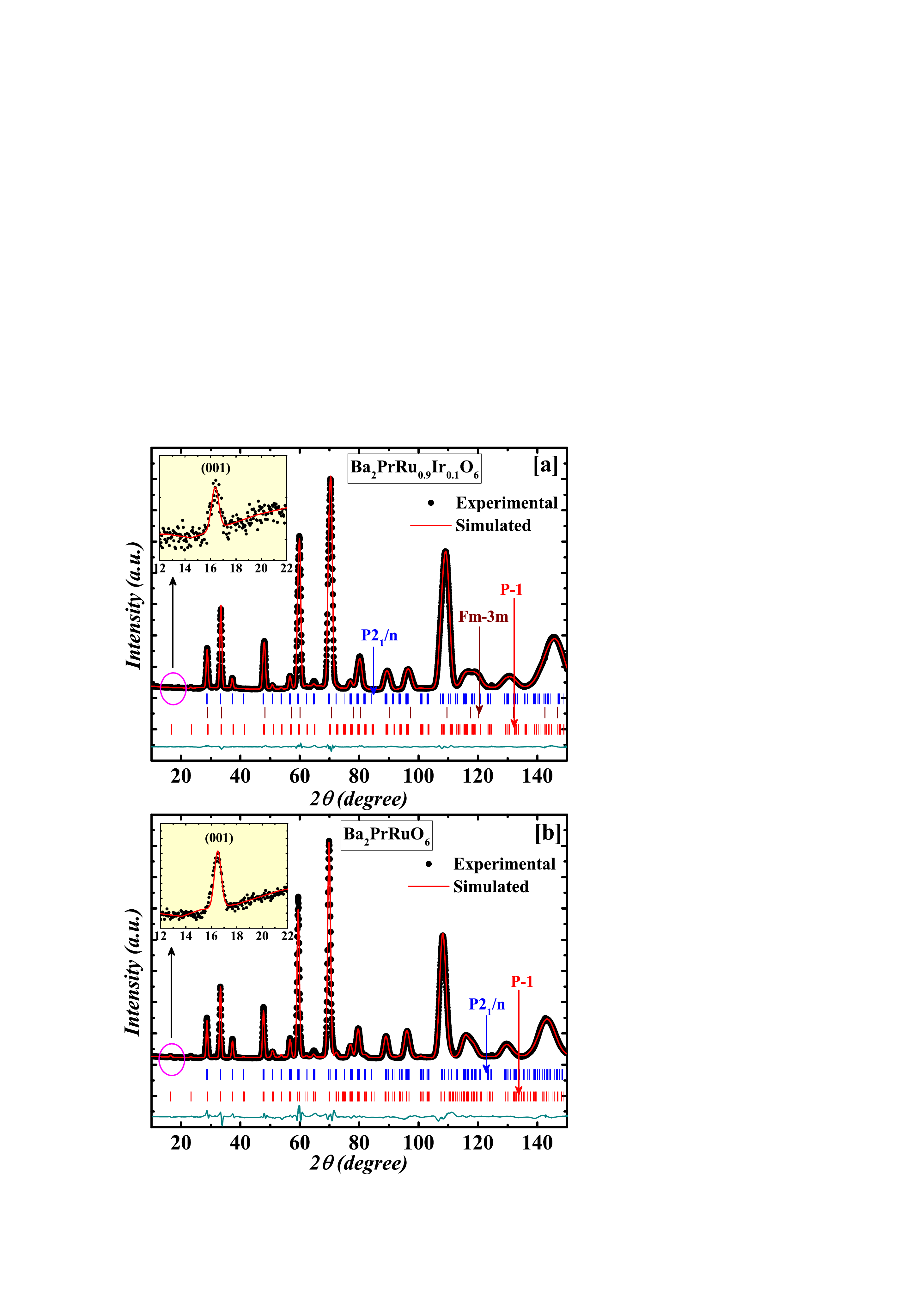}
\caption {(a) Rietveld refinement of powder neutron diffraction data recorded on D20 at 2 K.  Black points indicate the experimental data points, and the red solid line is the simulated curve. The structural Bragg reflections are indicated by blue and maroon vertical ticks for the monoclinic and cubic phases, respectively, and red ticks mark the magnetic Bragg reflections from the monoclinic phase. Inset: the $(001)$ magnetic reflection together with fitted peak. (b) shows the same for Ba$_2$PrRuO$_6$ with blue and red ticks for the structural (monoclinic) and magnetic Bragg reflections.}
\label{PND}
\end{figure}

\begin{figure}[t]
\includegraphics[width = 8 cm, trim={100 250 300 150}]{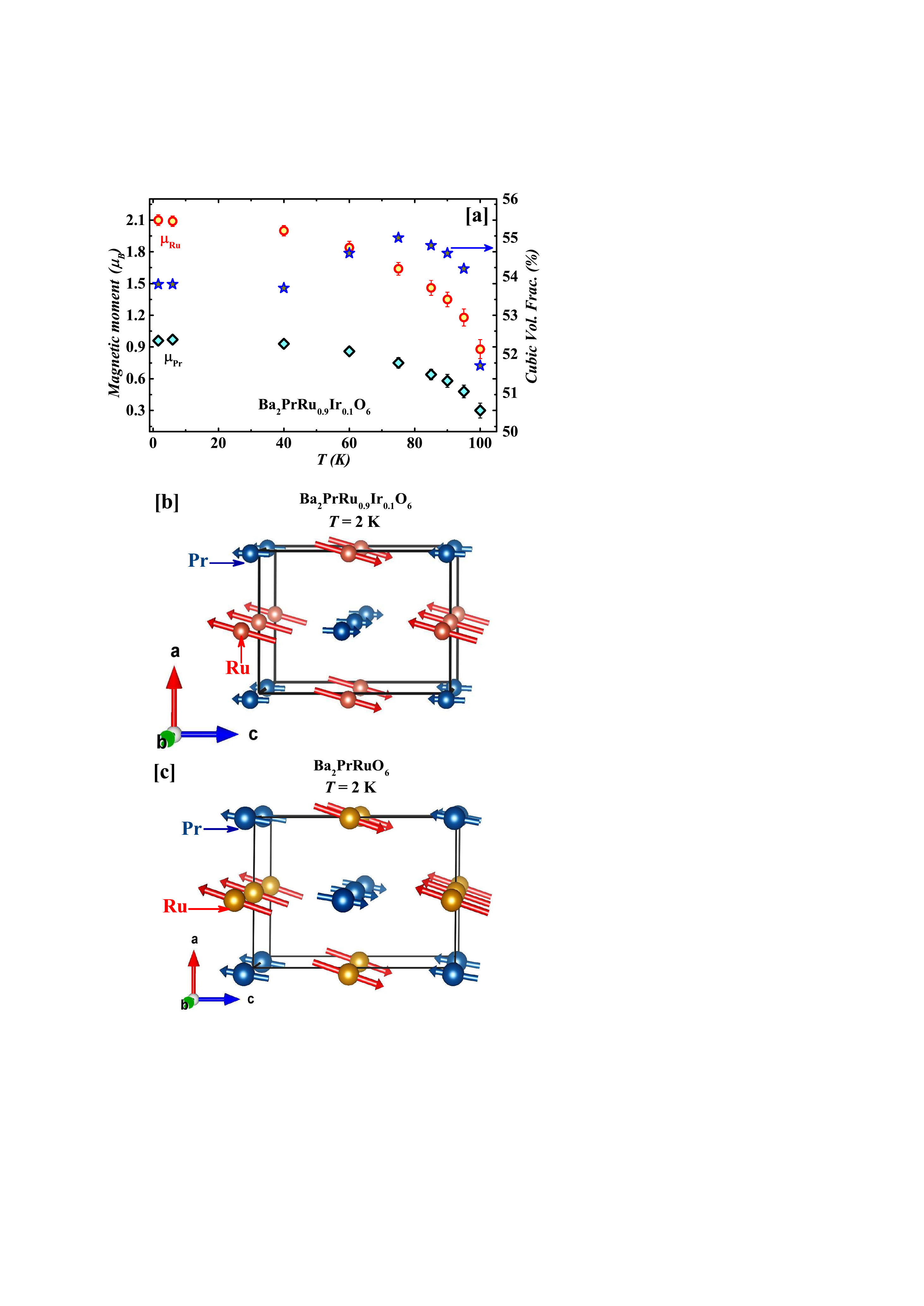}
\caption {(a) Shows the temperature dependence of Pr and Ru moment of the monoclinic phase in Ba$_2$PrRu$_{0.9}$Ir$_{0.1}$O$_6$. Blue stars indicate the volume fraction of the cubic phase (paramagnetic) with temperature. (b) and (c) depict the alignment of Pr and Ru spins in a magnetic unit cell of Ba$_2$PrRu$_{0.9}$Ir$_{0.1}$O$_6$ and Ba$_2$PrRuO$_6$ respectively.}
\label{magnetic structure}
\end{figure}

\subsubsection{Magnetic structure}
The magnetic susceptibility and structural studies clearly reveal the first order valence phase transition in  Ba$_2$PrRu$_{0.9}$Ir$_{0.1}$O$_6$ below approximately 200 K.  In Fig.~4 we present neutron diffraction results on the magnetic structure of Ba$_2$PrRu$_{0.9}$Ir$_{0.1}$O$_6$. The appearance of Bragg peaks at positions where no intensity from the structure of the two phases is present can be related to the long range order of magnetic moments. A rapid scan taken on cooling locates the onset of magnetic order at $T_{\rm N} \simeq 110$ K. The positions of the magnetic peaks are compatible with a propagation vector $\textbf{q} = (0,0,0)$ of the monoclinic phase. Magnetic symmetry analysis was performed with the Basireps software~\cite{basireps, ritter} to determine the allowed irreducible representations (irreps) for the Pr sublattice occupying the Wyckoff position $2c$ and the Ru sublattice at $2d$. There are only two allowed irreps, the same for both sites, which link the symmetry-related atomic positions antiferromagnetically in the $a$ and $c$ directions, and ferromagnetically in the $b$-direction, or vice versa. Only the first coupling scheme can account for the magnetic peak intensities, corresponding to the type-I antiferromagnetic structure (magnetic space-group $P\overline{1}$) found previously~\cite{izumi, parkinson} for Ba$_2$PrRuO$_6$. The FullProf software~\cite{FULLprof} was used to fit the 2 K diffraction data, and a stable refinement [see Fig.~4(a)] was achieved assuming the presence of two structural phases and one magnetic phase (the monoclinic structural phase, which orders magnetically, and the cubic structural phase for which no magnetic ordering was detected in our neutron diffraction study at 2 K). This has to be compared to the case of Ba$_2$PrRuO$_6$ for which a good refinement could be achieved considering only one monoclinic structural phase and one magnetic phase, as displayed in Fig.~4(b).
\par
Both the Pr and Ru sites carry a magnetic moment which is preferentially aligned along the $c$-axis.  Figure~5(a) shows the temperature dependence of the refined Pr and Ru moments and of the volume fraction of the cubic phase of Ba$_2$PrRu$_{0.9}$Ir$_{0.1}$O$_6$ as determined from the three-phase refinement of the D20 data. As the magnetic form factor of the Ru$^{5+}$ ion is not listed in the International Tables of Crystallography an empirical magnetic from factor which had been determined previously~\cite{parkinson1} has been used. The plot shows that at about 100 K, where the volume fraction of the cubic phase starts to decrease rapidly, the magnetic moments of Pr and Ru have decreased to nearly zero. As noted above, the magnetic Pr and Ru sublattices couple ferromagnetically in the $ab$ plane and these ferromagnetic layers are aligned antiferromagnetically along the $c$ axis as shown in Fig.~5(b).
\par
The existence of a magnetic component along the $a$-direction can be inferred from the presence of the purely magnetic $(001)$ peak. see the insets of Fig.~4(a) and (b). Because of the weakness of this peak, a refinement of the purely magnetic diffraction was undertaken for Ba$_2$PrRu$_{0.9}$Ir$_{0.1}$O$_6$ using a difference data set 2 K $-$ 130 K where the scale factor was fixed to the value obtained for the monoclinic phase fraction at 2 K. A refinement which allowed the presence of a magnetic component in $a$-direction on both sublattices did not converge as the Pr component fluctuates around zero. We therefore fixed the $a$ component for Pr to zero leading to a stable refinement with $\mu_{\rm Pr}^c$ = 1.10(1) $\mu_{\rm B}$, $\mu_{\rm Ru}^a$ = 0.55(3) $\mu_{\rm B}$ and $\mu_{\rm Ru}^c$ = 2.02(1) $\mu_{\rm B}$. A similar refinement of a difference data set 2 K $–$ 150 K was performed for the parent compound Ba$_2$PrRuO$_6$. This time, the presence of a magnetic component in the $a$-direction could be ascertained for Pr as well as for Ru: $\mu_{\rm Pr}^a$ = 0.26(9) $\mu_{\rm B}$, $\mu_{\rm Pr}^c$ = 1.09(1) $\mu_{\rm B}$, $\mu_{\rm Ru}^a$ = 0.30(9) $\mu_{\rm B}$ and $\mu_{\rm Ru}^c$ = 1.88(1) $\mu_{\rm B}$ [Fig.~5(c)]. We note that for Ba$_2$PrRuO$_6$, previous reports mention only a $c$-axis component for both sublattices.~\cite{izumi, parkinson}

\begin{figure}
\includegraphics[width = 8 cm, trim={0 0 150 450},clip]{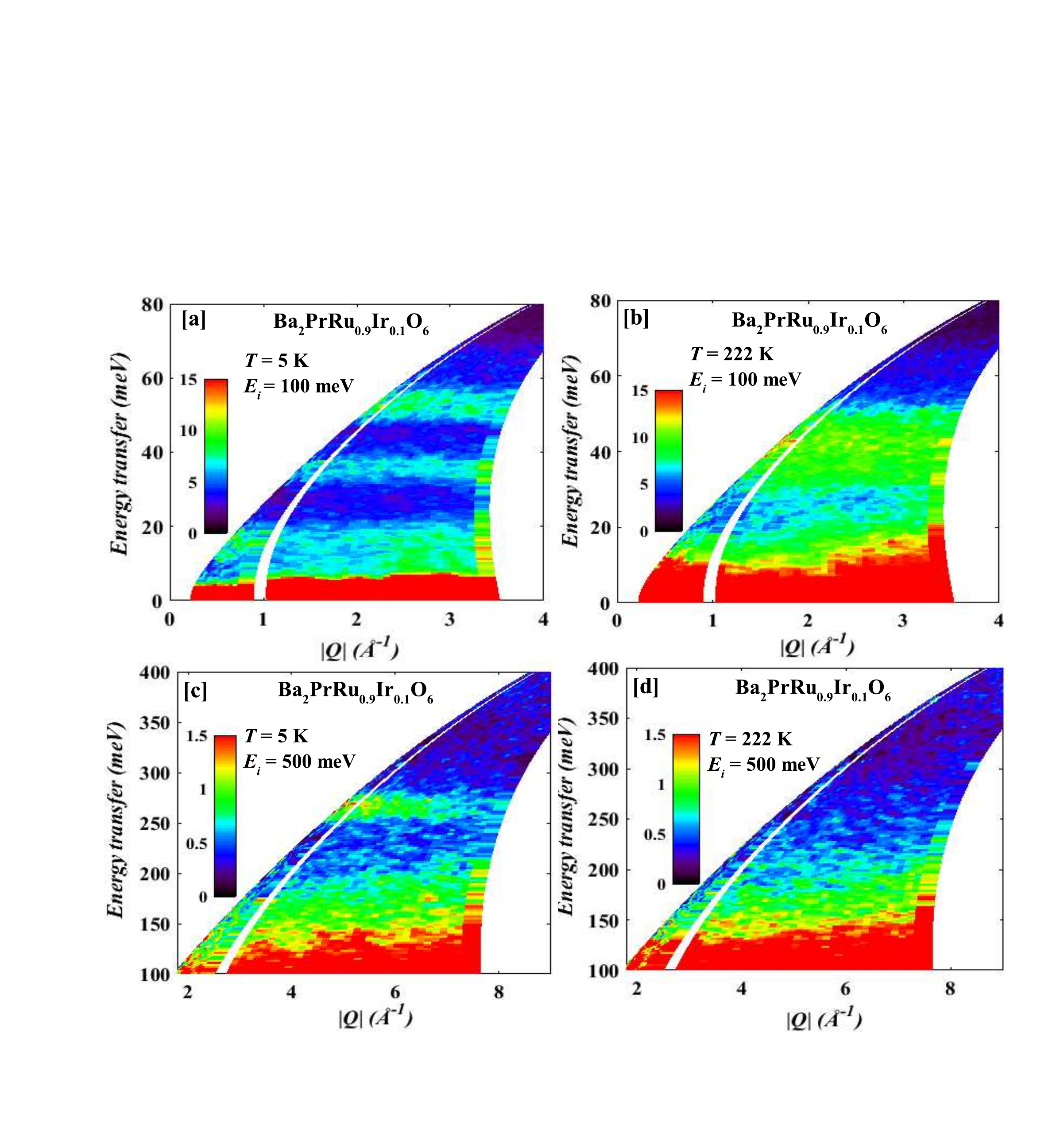}
\caption {Neutron scattering spectra of Ba$_2$PrRu$_{0.9}$Ir$_{0.1}$O$_6$. (a) and (b) show color-coded intensity maps as a function of energy transfer ($\hbar\omega$) and momentum transfer ($|\textbf{Q}|$) at 5 K and 222 K, respectively, measured with an incident energy $E_{\rm i} = 100$ meV. (c) and (d) display the same plots for $E_{\rm i} = 500$ meV.}
\label{INS}
\end{figure}

\subsection{Inelastic neutron scattering study}
Figure~6 presents color-coded INS intensity maps as a function of energy transfer ($\hbar\omega$) and momentum transfer ($Q= |\textbf{Q}|$) for $E_{\rm i} = 100$ meV and 500 meV at $T = 5$ and 222 K. The $E_{\rm i} = 100$ meV spectrum at 5 K, Fig.~6(a), shows four dispersionless bands of scattering at $\hbar\omega = 15$, 35, 55 and 68 meV, while the $E_{\rm i} = 500$ meV spectrum at 5 K, Fig.~6(c), contains a dispersionless band at $\hbar\omega = 264$ meV. The 264 meV excitation is absent at 222 K, Fig.~6(d), while the excitations between 35 and 68 meV transform from four well-defined peaks into a very broad (in energy) signal centered near 45 meV at 222 K, Fig.~6(b). The lack of dispersion with $Q$, and the reduction in intensity with $Q$ consistent with a magnetic form factor, implies that these signals arise from localized crystalline electric field (CEF) transitions of the Pr ions.

Figures~7(a)--(d) show spectra measured at different $E_{\rm i}$ from 60 to 500 meV, averaged over a small range of $Q$ in the low $Q$ regime. Data for both $T = 5$ K and 222 K are included. These spectra confirm that at 222 K (also at 130 K not shown here) there is a single broad peak centred near 45 meV, whereas at 5 K four well-defined peaks appear at around 15,  35, 55 and 68 meV (also observed at 65 K, not shown here). In the 500 meV spectra, Fig.~7(d), the peak at $\hbar\omega = 264$ meV in the 5 K spectrum is completely absent from the 222 K spectrum.

We identify the 264 meV peak as a CEF transition within the  $J = 5/2$ ground state level of Pr$^{4+}$ in the cubic phase, and the four low-energy peaks as CEF transitions of Pr$^{3+}$ in the monoclinic phase. This interpretation follows from the observations described earlier, that the monoclinic phase is present both above and below the valence transition temperature $T_{\rm V} \simeq 170$ K, whereas the cubic phase is only present at $T < T_{\rm V}$.

Further evidence to support the assignment of the 264 meV peak comes from a comparison with the the INS spectrum of BaPrO$_3$, which contains a peak at 255 meV.~\cite{kern} In BaPrO$_3$, the Pr$^{4+}$ ions have a slightly distorted octahedral coordination very similar to that in the monoclinic phase of  Ba$_2$PrRu$_{0.9}$Ir$_{0.1}$O$_6$. The 255 meV in BaPrO$_3$ was found to originate from the $\Gamma_7$ to $\Gamma_8$ transitions within the $J = 5/2$ level.~\cite{kern} Furthermore, we have observed a similar CEF excitation at 276 meV in Ba$_2$PrIrO$_6$,~\cite{winfred} also attributed to the same group of $\Gamma_7$ to $\Gamma_8$ transitions of Pr$^{4+}$.

\begin{figure}
\includegraphics[width = 8 cm,trim={20 0 50 50}]{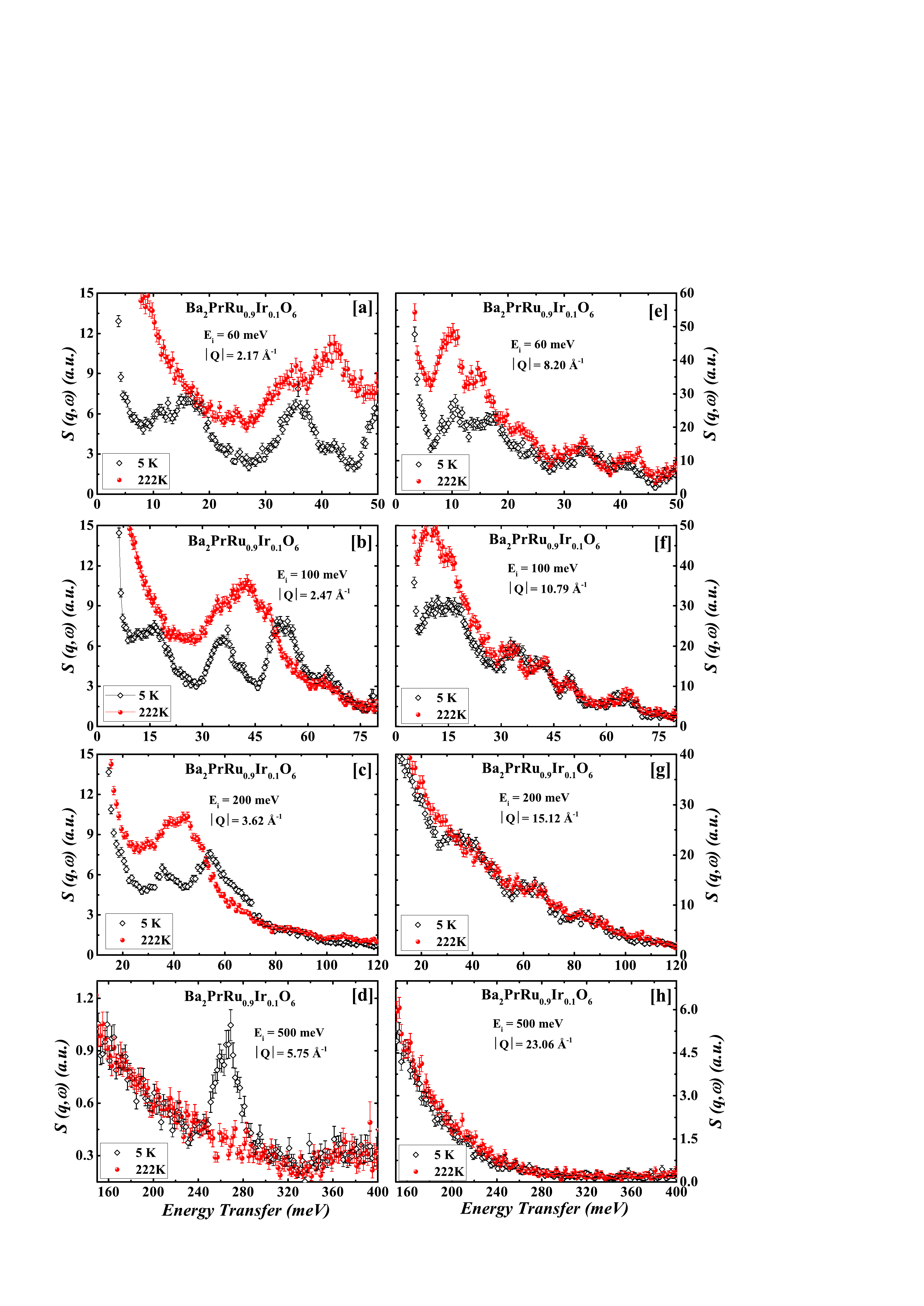}
\caption {$Q$-averaged INS intensity as a function of energy transfer at different $E_i$ (= 60, 100, 200 and 500 meV). Left and right panels show the low-$Q$ and high-$Q$ regimes, respectively.}
\label{INS1}
\end{figure}

Figures~7(e)--(h) show spectra for averaged $Q$ values in the high $Q$ regime. In this regime the spectra are dominated by phonon and muti-phonon scattering. As expected, the low energy part of the spectra increase in intensity with temperature due to thermal population of phonons. There are also some small changes in the peak positions and widths between 5 K and 222 K which are possibly, but not conclusively, associated with the structural transition.

\begin{figure}
\includegraphics[width = 8 cm,trim={50 0 150 600}]{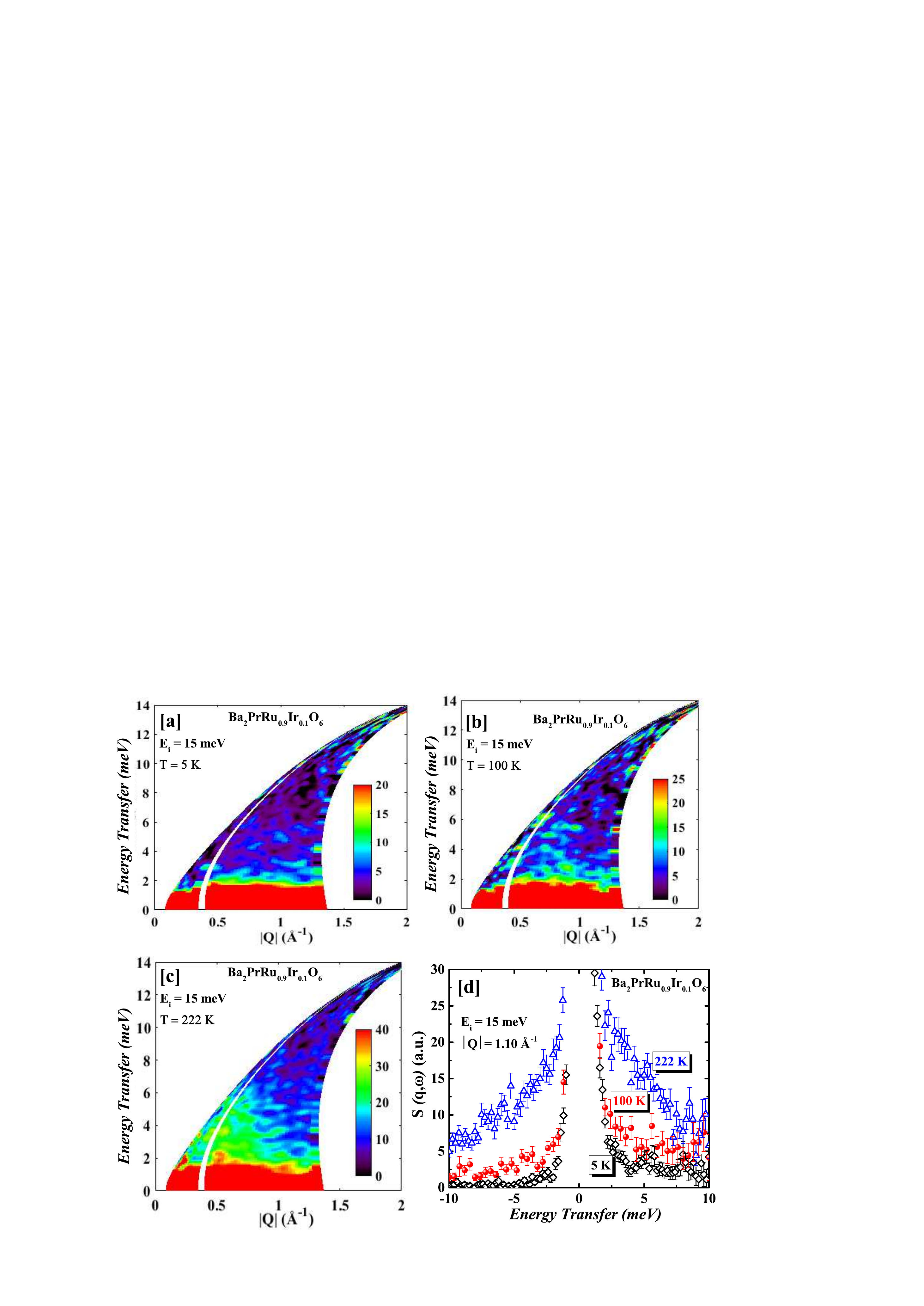}
\caption {INS spectrum of Ba$_2$PrRu$_{0.9}$Ir$_{0.1}$O$_6$ measured on HET with $E_{\rm i} = 15$ meV. (a), (b) and (c) show the color-coded intensity maps as a function of energy transfer and $Q$ recorded at 5 K, 100 K and 222 K. (d) is the $Q$-averaged scattering at the same temperatures as a function of energy transfer.}
\label{INS2}
\end{figure}

Figure~8 shows INS spectra measured with $E_{\rm i}$=15 meV. The spectrum at $T = 100$ K shows a small increase in scattering compared with that at 5 K, but at 222 K the spectrum displays strong additional quasielastic scattering. As discussed earlier, 222 K is above both the magnetic ordering transition ($T_{\rm N} \simeq 110$ K) and the valence transition ($T_{\rm V} \simeq 170$ K), and at 222 K there exists only a single structural phase. We attribute the quasielastic signal observed at 222 K, therefore, to paramagnetic scattering from the Pr$^{3+}$ and Ru$^{5+}$ ions in the monoclinic phase. The reduction in scattering on cooling, first to 100 K and then 5 K, could be accounted for by the freezing of dynamic magnetic fluctuations due to magnetic order, and a reduction in the volume fraction of the monoclinic phase below $T_{\rm V}$. The significant widths of both the quasielastic scattering and the inelastic peaks observed at 222 K may be due to residual magnetic correlations above $T_{\rm N}$ and/or to lifetime damping due to magnetoelastic effects or valence fluctuations associated with the incipient valence transition. We have also measured pure Ba$_2$PrRuO$_6$ with the same $E_{\rm i}$ values and we find very similar changes in the excitations when going from 147 K to 7 K (data not shown here).

We now present a more quantitative analysis of the CEF splitting of the Pr ions in order to provide further support for our interpretation of the spectra. We consider first the Pr$^{4+}$ ions in the cubic phase. The $4f^1$ configuration of Pr$^{4+}$ is split by the spin--orbit interaction ($\zeta = 107.2$ meV) into two levels, $^2F_{5/2}$ ($J = 5/2$) and $^2F_{7/2}$ ($J = 7/2$) with 6-fold and 8-fold degeneracy ($= 2J +1$), respectively. The degeneracy is lifted by the CEF from the neighbouring ions, which has octahedral point symmetry ($m\overline{3}m$ or $O_h$) and is described by the Hamiltonian
\begin{equation}
{\mathcal H}_{\rm{CEF}}= B_0^4\mbox{\Large [}C_0^{(4)}\pm \frac{\surd{70}}{14}C_4^{(4)}\mbox{\Large ]} + B_0^6\mbox{\Large [}C_0^{(6)}+\mp \frac{\surd{14}}{2}C_4^{(6)}\mbox{\Large ]},
\label{CEF-Hamiltonian}
\end{equation}
where $B_q^k$ are phenomenological CEF parameters, and $C_q^{(k)}(\theta, \phi) = \sqrt{\frac{4\pi}{(2k+1)}}Y_{k,q}(\theta, \phi)$ are tensor operators as defined in Ref.~\onlinecite{Wybourne}, with Y$_{k,q}(\theta$, $\phi$) the spherical harmonics.  The relation between the CEF parameters of the tensor operators (i.e.~the $B_q^k$ used here) and the commonly-used Stevens operator equivalents can be found in Ref.~\onlinecite{Kassman}, although we caution that the Stevens operator method is not accurate for the present system because the CEF splitting is comparable to the spin-orbit splitting.

The cubic CEF described by (\ref{CEF-Hamiltonian})  splits the ground state $^2F_{5/2}$  level of Pr$^{4+}$ into a doublet of $\Gamma_7$ symmetry and a quartet of $\Gamma_8$ symmetry, and splits the $^2F_{7/2}$ upper level into two doublets ($\Gamma_6'$, and $\Gamma_7'$) and a quartet ($\Gamma_8'$). The prime notation is used here to distinguish the states belonging to $^2F_{7/2}$ from those of  $^2F_{5/2}$.

In Fig.~9 we show the INS spectra recorded with the two highest $E_{\rm i}$ values. The experimental spectra are magnified to emphasize the weak features, and are plotted for a fixed average scattering angle $\phi$ rather than for a fixed average $Q$. The lines are simulated spectra calculated from eq.~(\ref{CEF-Hamiltonian}) with the program SPECTRE~\cite{SPECTER} and the cross-section formulae summarized in Ref.~\onlinecite{Osborn1991}. The ratio $B_0^6/B_0^4$ was initially fixed by the point charge model for nearest-neighbours only, and $B_0^4$ was varied to put the $\Gamma_8$ level at 264 meV.  As well as the $\Gamma_7$ to $\Gamma_8$ transition at 264 meV, the calculation predicted an inter-level transition from $\Gamma_7$ to $\Gamma_7'$ at around 420 meV. This coincides with a small peak in the data centered on about 400 meV, see Fig.~9(a). A similar feature was observed in the INS spectrum of Ba$_2$PrIrO$_6$ at about 370 meV.\cite{winfred} Subsequently, $B_0^6$ was allowed to vary slightly. The simulations shown in Figs.~9(a) and (b) are calculated with CEF parameters (in Wybourne tensor notation) $B_0^4 = 971$ meV and $B_0^6 = 60$ meV, and Fig.~9(c) displays the full CEF level scheme of Pr$^{4+}$ for these parameters.

The point symmetry for the Pr$^{3+}$ ions ($4f^2$) in the monoclinic phase is triclinic ($\overline{1}$ or $C_i$). In such a low symmetry the CEF Hamiltonian required 15 parameters, and the ground state $^3H_4$ level is maximally split into 9 singlets. Given that we have observed only four CEF transitions at $T=5$ K, a quantitative analysis of the Pr$^{3+}$ data is not feasible. To gain a rough estimate of the contribution from Pr$^{3+}$ to the high energy spectra we estimated the Pr$^{3+}$ CEF parameters by scaling those for Pr$^{4+}$ using relations applicable to the point charge model. A simulation of the Pr$^{3+}$ spectrum obtained this way is shown on Figs.~9(a) and (b). We assumed Pr$^{4+}$ and Pr$^{3+}$ to be presented in the ratio 0.58:0.42, as observed experimentally.  The largest peak from Pr$^{3+}$ in the high energy range is found at 330 meV, and corresponds to an inter-level transition from $^3H_4$ to  $^3H_5$. This peak is nearly an order of magnitude smaller than the 264 meV peak, and could contribute to the small peak observed in the data at around 400 meV. More importantly, the simulation confirms that the 264 meV peak is from Pr$^{4+}$, because there is no peak of comparable strength from Pr$^{3+}$ in this energy range.

Finally, we remark that in PrO$_2$, which also has Pr$^{4+}$ in a cubic CEF environment, but in 8-fold rather than 6-fold coordination, the CEF-split levels are reversed, with $\Gamma_8$ as the ground state and $\Gamma_7$ as the excited state. In addition to the $\Gamma_8$ to $\Gamma_7$ transition near 130 meV and inter-level transitions at higher energies, the INS spectrum of PrO$_2$ contains broad magnetic peaks near 30 meV and 160 meV.~\cite{PrO2} These broad continua are attributed to vibronic modes due to the dynamic Jahn--Teller effect. Such excitations occur in PrO$_2$ because of the large orbital degeneracy of the $\Gamma_8$ ground state, and so are not expected to be observed for Pr$^{4+}$ in octahedral coordination, as in Ba$_2$PrRu$_{0.9}$Ir$_{0.1}$O$_6$.

\par

\begin{figure}[t]
\centering
\includegraphics[width =8 cm, trim={50 100 150 0}]{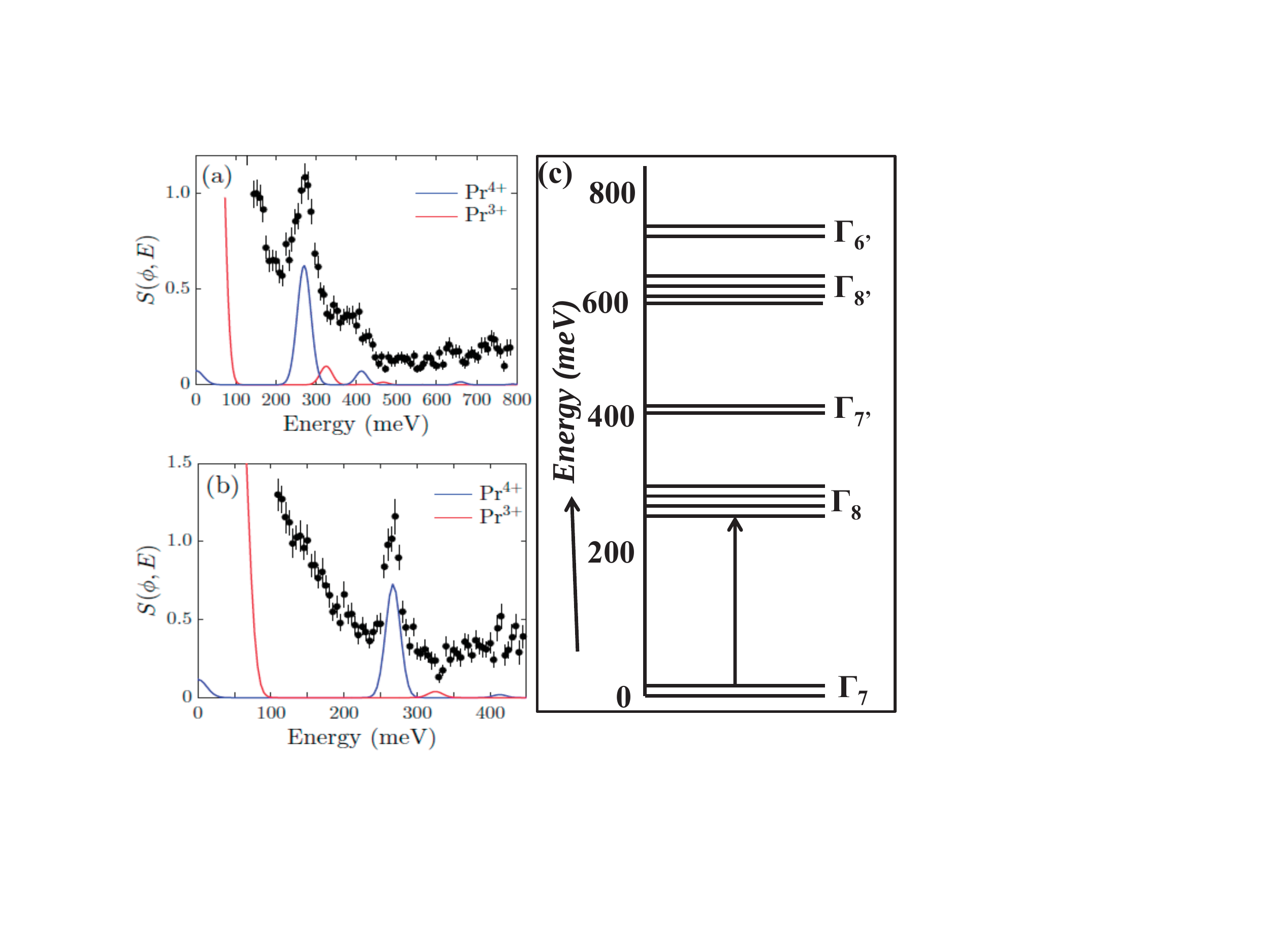}
\caption {INS spectra of Ba$_2$PrRu$_{0.9}$Ir$_{0.1}$O$_6$ at fixed scattering angle $\phi$ as a function of energy transfer at 5 K, for incident energies of (a) $E_{\rm i} = 900$ meV,  averaged from $\phi =3$ to $10^\circ$ ($\langle \phi \rangle = 6.5^\circ$), and (b) $E_{\rm i} = 500$ meV, averaged from $\phi =3$ to $12^\circ$ ($\langle \phi \rangle = 7.5^\circ$). The solid lines show the simulated INS spectra for Pr$^{4+}$ (blue) and Pr$^{3+}$ (red) calculated from the cubic CEF model described in the text.  (c) Energy level scheme for the $J=5/2$ and $J=7/2$ levels of Pr$^{4+}$ calculated from the CEF model. }
\label{CEF_simulation}
\end{figure}


\begin{figure}[t]
\centering
\includegraphics[width = 8 cm,trim={20 150 100 350}]{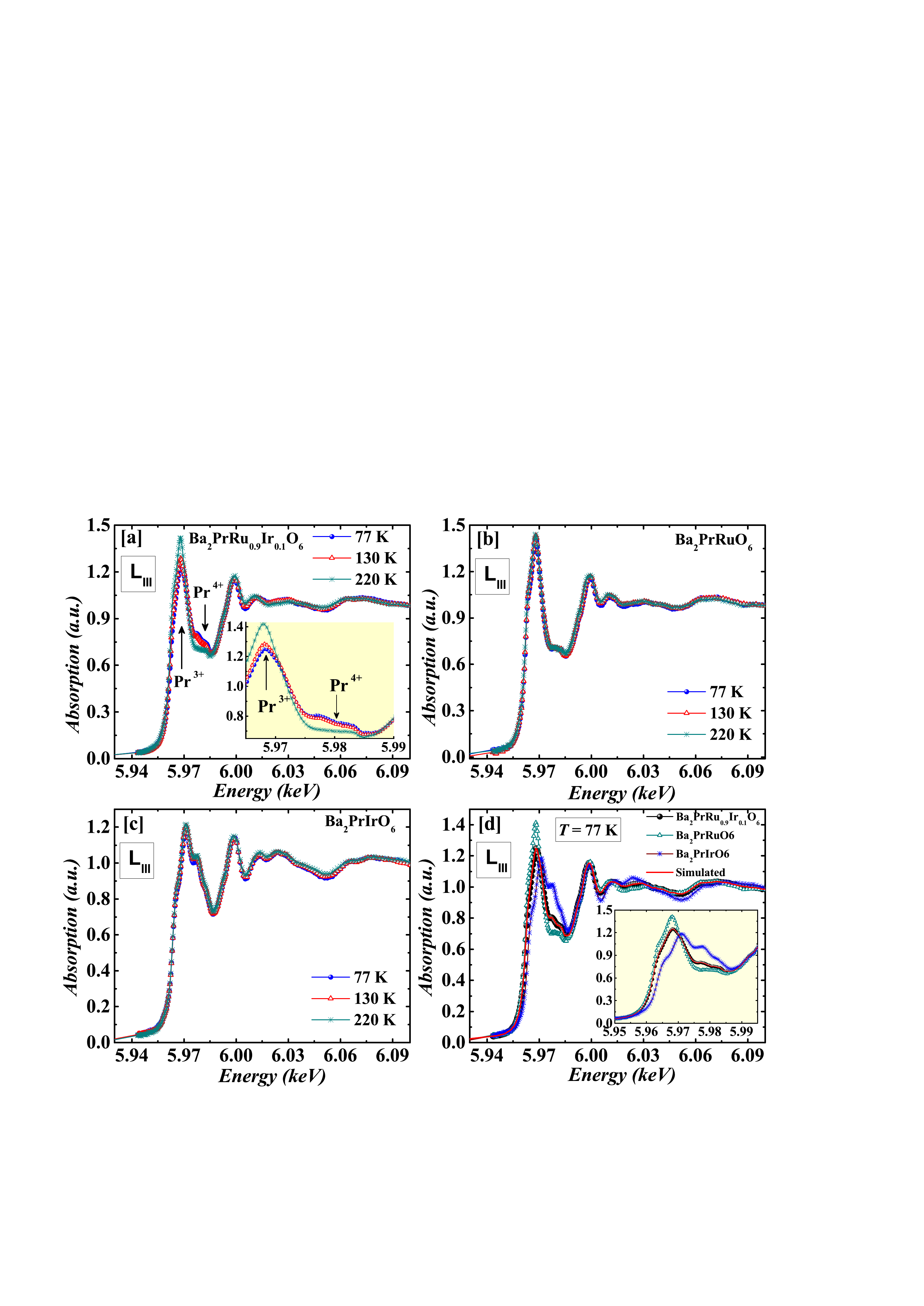}
\caption {XANES spectra at various temperatures are depicted for (a) Ba$_2$PrRu$_{0.9}$Ir$_{0.1}$O$_6$, (b) Ba$_2$PrRuO$_6$, and (c) Ba$_2$PrIrO$_6$. The inset of (a) shows an enlarged view of the anomalies near 5.966 keV and 5.979 keV. (d) shows the calculated spectra of Ba$_2$PrRu$_{0.9}$Ir$_{0.1}$O$_6$ from the observed spectra of Ba$_2$PrRuO$_6$ and Ba$_2$PrIrO$_6$ at 77K.}
\label{XANES}
\end{figure}
\par

\subsection{X-ray absorption spectroscopy}
We have investigated the Pr $L_{3}$-edge x-ray absorption near-edge structure (XANES) spectrum of Ba$_2$PrRu$_{1-x}$Ir$_x$O$_6$ with $x=0.1$ in order to obtain direct information on the valence states of the Pr ion below and above the transition. As a reference, we have also measured the compounds with $x=0$ and $x=1$. The Pr $L_{3}$-edge measurements were carried out as a  function of temperature, above and below the first-order phase transition.  Figures~10(a) to (c) show Pr-$L_{3}$ measurements at 77 K, 130 K and 220 K (data were collected during a heating cycle) for  Ba$_2$PrRu$_{1-x}$Ir$_x$O$_6$ with $x=0.1$, 0 and 1, respectively.

It can be seen in Figs.~10(b) and (c) that the spectra of Ba$_2$Pr$^{3+}$RuO$_6$ and Ba$_2$Pr$^{4+}$IrO$_6$ do not show any temperature dependence, whereas those of Ba$_2$PrRu$_{0.9}$Ir$_{0.1}$O$_6$ exhibit a strong temperature dependence near 5.966 keV ($L_{3}$ of Pr$^{3+}$) and 5.979 keV ($L_{3}$ of Pr$^{4+}$). The peak at 5.966 keV arises due to the excitation from $2p_{3/2}$ to $4f^2 5d^*$, which represents Pr$^{3+}$ ions. The other feature near 5.979 keV emerges from $2p_{3/2}$ to $4f^2 \underline{L} 5d^*$ ($\underline{L}$ being the ligand hole) and $4f^1 5d^*$ excitations which represent Pr$^{4+}$ ions.~\cite{bianconi, fujishiro} The intensity near 5.979 keV is higher at 77 K and 130 K than at 220 K, which indicates directly that at low temperatures the Pr valence changes from 3+ to 4+. This is also evidenced from the decrease in the intensity of the 5.966 eV peak at 130 K and 77 K. The inset of Fig.~10(a) shows clearly this temperature dependence. A very similar behaviour was also observed for the Pr $L_2$-edges (data not shown here).

To estimate the volume fraction of the Pr$^{3+}$ and Pr$^{4+}$ phases at 77 K in Ba$_2$PrRu$_{0.9}$Ir$_{0.1}$O$_6$, we have added the intensity of the two end members (i.e. $x=0$ and 1), $y$Ru + $(1-y)$Ir, and compared the calculated intensity with the observed intensity of the $x=0.1$ composition. The result is shown in Fig.~10(d). The observed and calculated intensities agree well for $y\approx 0.36$. This indicates that the volume fraction of the Pr$^{4+}$ phase, i.e. (1-y), is approximately 64\% at 77 K, which is in agreement with that estimated from the neutron diffraction data discussed above.


\section{Conclusions}
We have investigated Ba$_2$PrRu$_{1-x}$Ir$_x$O$_6$ ($x=0.1$) with neutron diffraction, inelastic neutron scattering and x-ray absorption measurements, as well as susceptibility and heat capacity, and compared the results with $x=0$ and 1. All our results confirm the first order nature of a valence phase transition of Pr below 200 K in Ba$_2$PrRu$_{1-x}$Ir$_x$O$_6$ ($x=0.1$).

Our high resolution neutron diffraction study at 300 K reveals that the $x=0.1$ sample is single phase and crystallized in the monoclinic phase, whereas below 170 K two phases coexist, the monoclinic phase (containing Pr$^{3+}$) and a cubic phase (containing Pr$^{4+}$). We have estimated the volume faction of these two phases and found 58\% for the cubic and 42\% for the monoclinic phase at 5 K.

Furthermore, our neutron diffraction study clearly shows a long range magnetic ordering of both Pr and Ru moments below 100 K  in the monoclinic phase. Analysis of the $x=0.1$ data at 2 K revealed that the Pr moments point along the $c$ axis, but the Ru moments have a minority $a$ axis component in addition to a majority $c$-axis component.  On the other hand for $x=0$ we have found that both Pr and Ru moments are tilted away from the $c$-axis.

Our high energy inelastic neutron scattering study on $x=0.1$ reveals an excitation at 264 meV which is present at 5 K but absent at 222 K, consistent with a change of the valence state of some of the Pr ions from Pr$^{4+}$ at low temperature to Pr$^{3+}$ above 200 K, corresponding to a transition from cubic to monoclinic structures.

Finally, we have observed clear evidence for the Pr valence transition in our x-ray absorption measurements  on $x=0.1$, in contrast to $x=0$ and 1 for which no valence transition was observed.
\\

\section*{ACKNOWLEDGEMENTS}
We thank B.D. Rainford for interesting discussion, D. Visser for participanting in the x-ray experiment and Y. Izumiyama for help in the samples preparation. DTA and ADH would like to thank CMPC-STFC, grant number CMPC-09108, for financial support. DTA would like to thank  JSPS for the invitation fellowship. J.S. would like to thank the European Union’s Horizon 2020 research and innovation programme under the Marie Skłodowska-Curie grant agreement (GA) No 665593 awarded to the Science and Technology Facilities Council. The authors would like to thank Diamond Light Source for beamtime on B18 (proposal No SP63810-1) and ISIS Facility for beam time on HRPD and HET.

\par

%
\end{document}